\documentclass[letterpaper,showpacs,aps,superscriptaddress]{revtex4-2}

\usepackage[utf8]{inputenc}
\usepackage[english]{babel}

\usepackage{latexsym}
\usepackage{amssymb}
\usepackage{amsmath}
\usepackage{bm} 

\usepackage{epsfig}
\usepackage{graphicx}
\usepackage{float}

\usepackage{color}

\usepackage{tensor} 

\usepackage[colorlinks=true,allcolors=blue]{hyperref}

\usepackage{comment}

\begin{document}


\title{$\ell$-Proca stars}

\author{Claudio Lazarte}
\email{claudio.lazarte@uv.es}
\affiliation{Instituto de Ciencias Nucleares, Universidad Nacional
Aut\'onoma de M\'exico, Circuito Exterior C.U, A.P. 70-543, M\'exico D.F. 04510, M\'exico}
\affiliation{Departamento de Astronom\'ia y Astrof\'isica, Universitat de Val\`encia,
Dr. Moliner 50, 46100, Burjassot (Val\`encia), Spain}

\author{Miguel Alcubierre}
\email{malcubi@nucleares.unam.mx}
\affiliation{Instituto de Ciencias Nucleares, Universidad Nacional
Aut\'onoma de M\'exico, Circuito Exterior C.U, A.P. 70-543, M\'exico D.F. 04510, M\'exico}




\begin{abstract} 
Initially applied to the scalar case, we extend the applicability of the multi-field generalization with angular momentum of bosonic stars to the vector case, in order to obtain new configurations that generalize the one-field spherical Proca stars. These new objects, which we call $\ell$-Proca stars, arise as stationary and spherically symmetric bosonic stars solutions of the Einstein-(multi)Proca system, whose matter content is formed by an arbitrary odd number of $2\ell+1$ of complex Proca fields with the same mass, time-frequency, radial profile and angular momentum number $\ell$. We analyze the system of constraint and evolution radial equations for the matter content to show the consistency of our proposal, and obtain numerically the ground states of these new solutions for the first few values of $\ell$ using spectral methods. 
\end{abstract}




\maketitle


\section{Introduction}
\label{sec:intro}

Proca stars~\cite{BRITO2016291} are stationary classical solutions of the Einstein--(complex)Proca system describing self-gravitating, everywhere regular, horizonless configurations for a single massive complex vector field. They are classified as bosonic stars, a class that also includes their scalar cousins, the usual (scalar) boson stars~\cite{Kaup68,Ruffini69}.  Although Proca stars remain theoretical at present, they arise as simple models of exotic compact objects with dynamical robustness~\cite{Sanchis_Gual_2017,Sanchis_Gual_2019fae, DiGiovanni_2020,herdeiro2023} and a non-fine-tuned formation mechanism \cite{DiGiovanni2018,Sanchis_Gual_2019fae,herdeiro2023}, which makes them dynamically viable candidates to deal with the problem of degeneracy in the interpretation of strong gravity data against the black hole paradigm. Indeed, to address this issue, studies of the emission of gravitational radiation from Proca-star mergers have been performed \cite{Sanchis_Gual_2019hop,Sanchis-Gual_2022}, allowing systematic searches for exotic compact object mergers in the detected gravitational wave events \cite{Bustillo2021,bustillo2023svb,Bustillo2023inp}. These studies have found, for example, that the gravitational wave signal GW190521~\cite{LIGO:gw190521} is in fact consistent with a head-on collision of Proca stars, an interpretation statistically slightly preferred over the binary black hole hypothesis, paving the way toward population studies of more complex exotic compact objects as generalizations of Proca stars.

Standard bosonic stars can be generalized to multi-field configurations formed by $2\ell+1$  ($\ell \in \mathbb{N}_0$)  complex bosonic fields with the same mass, same time-frequency and same radial profile, sourcing a static and spherically symmetric spacetime. This is achieved by extending the internal symmetry group of their matter content from $U(1)$ to $U(2\ell+1)$. Thus, the internal group can hide not only the time dependency of each field but also the angular dependency given by the different spherical harmonics available for a fixed angular momentum number $\ell$, provided their radial-dependent profiles are the same for the $2\ell+1$ matter fields. This idea, inspired by a similar idea used first in the context of gravitational collapse by Olabarrieta {\em et al.}~\cite{Olabarrieta:2007di}, was applied in the case of (scalar) boson stars finding its respective generalization known as ``$\ell$-boson stars"~\cite{Alcubierre:2018} (recently a similar idea has also been applied in the case of the Einstein--Dirac system by Shi-Xian Sun {\em et al.} resulting in the so-called ``$k$-Dirac stars"~\cite{sun2023kappadirac}.). These configurations reduce to standard boson stars in the case of $\ell=0$, and with larger values of $\ell$ they result in more massive and more compact bosonic stars.  Further studies of $\ell$-boson stars have been done, such as for example the exploration of the parameter space of extreme configurations with large values of $\ell$~\cite{Alcubierre:2022}, or the study of the dynamics of head-on collisions and the extraction of the gravitational radiation~\cite{Jaramillo:2022hce}. Recently, a semiclassical gravity solution for a single  real massive quantum scalar field was shown to be a more natural way of obtaining $\ell$-boson star configurations, now obtained as self-gravitating distributions of identical spin 0 particles with the same definite energy and angular momentum, arising as a particular excitation of such a single quantum field~\cite{Alcubierre:2023}. This result shows that these kind of multi-field generalizations are well-motivated by the quantum nature of bosonic fields.

Stability analysis of $\ell$-boson stars was performed with dynamical time evolutions of both linear~\cite{Alcubierre:2021} and fully nonlinear spherical~\cite{Alcubierre:2019} perturbations, as well as nonlinear nonspherical perturbations~\cite{Jaramillo:2020}. The last study suggests the existence of nonspherical equilibrium configurations produced by perturbing $\ell$-boson stars, that were later shown to be part of a larger continuous family of multi-field and multi-frequency boson stars in~\cite{Sanchis-Gual:2021}. The same work \cite{Sanchis-Gual:2021} asserts that $\ell$-boson stars are the symmetry-enhanced points and unique stable configurations of this family, and conjectures that an analogous family for Proca stars should exist. In order to establish the first step towards the proof of that conjecture, the goal of our work here is to extend this well motivated generalization idea to the case of Proca stars, obtaining in this way the classical solutions for $\ell$-Proca stars. 

This article is organized as follows. In Section~\ref{sec:system} we propose an ansatz based on $2\ell+1$ complex Proca fields with the same mass, same radial profile, and fixed angular momentum number $\ell$, that solves the spherically symmetric Einstein--(multi)Proca system.  Employing the 3+1 formalism we also present the evolution and constraint equations for the Proca fields of this system, and the matter source terms coming from its stress-energy tensor. Section~\ref{sec:procastar} imposes the harmonic time-dependency on our previous system, thus obtaining static and spherically symmetric multi-field Proca stars with angular momentum, which we call $\ell$-Proca stars. We then derive the system of equations that satisfy the spacetime and matter constraints and prescribe the radial dependency of these configurations. In Section~\ref{sec:localsolutions} we analytically find the local solutions of $\ell$-Proca stars for small radius in order to determine the regularity conditions at the origin.  Section~\ref{sec:results} mentions the essential details of the multi-domain collocation spectral method that were numerically implemented to obtain our solutions, and presents our numerical results for different values of $\ell$. In particular, we show the radial profiles of the ground state configurations, and the characteristic curves for families of those solutions.  We conclude in Section~\ref{sec:CONCLUSIONS}. Throughout this paper we use the signature $(-,+,+,+)$ for the spacetime metric, and Plank units such that $G = c = \hbar = 1$.


\section{The Spherically Symmetric Einstein--(multi)Proca system}
\label{sec:system}

We consider an arbitrary number of complex Proca fields $(X_m)_\alpha$, with $m=1,...,N$, each of mass $\mu$ and minimally coupled to gravity. This system is described by the following action:
\begin{equation}\label{Action:E-(multi)Proca}
S = \int d^4x \sqrt{-g}\ \left( \frac{R}{16\pi} - \frac{1}{16\pi} \sum^{N}_{m=1} \Big[ \ (W_m)_{\mu\nu}(\bar{W}_m)^{\mu\nu} + 2 \mu^2 (X_m)_\mu (\bar{X}_i)^{\mu} \Big] \right) \; ,
\end{equation}
where the Proca field strength tensor is given by $(W_m)_{\mu\nu} = \nabla_\mu (X_m)_\nu - \nabla_\nu (X_m)_\mu$, and the complex conjugates are denoted by an overbar. Due to its multi-field character, we refer to this as the Einstein-(multi)Proca system. The equations of motion for the gravitational field are the Einstein equations:
\begin{equation}\label{eq:field:Einstein}
R_{\mu\nu}-\frac{1}{2} \: R\ g_{\mu\nu} = 8 \pi T_{\mu\nu},
\end{equation}
where the stress-energy tensor is given by:
\begin{align}\label{exp:tensor:stress-energy}
T_{\mu\nu} = \frac{1}{4\pi} \sum^N_{m=1} \Big\{ &  - (W_m)_{\lambda(\mu}\tensor{(\bar{W}_m)}{_\nu_)^\lambda} - \frac{g_{\mu\nu}}{4} \: (W_m)_{\alpha\beta}(\bar{W}_m)^{\alpha\beta} \nonumber \\
& + \mu^2 \left[ (X_m)_{(\mu}(\bar{X}_m)_{\nu)} - \frac{g_{\mu\nu}}{2} \: (X_m)_\lambda (\bar{X}_m)^\lambda \right] \Big\} \; .
\end{align}
The corresponding equations of motion for the matter fields are the Proca equations:
\begin{equation}\label{eq:field:Proca}
\nabla_\mu (W_m)^{\mu\nu} - \mu^2 (X_m)^\nu = 0 \; .
\end{equation}

Although each massive vector field $(X_m)_\mu$ does not possess gauge freedom, it can be shown from Eq.~\eqref{eq:field:Proca} that it must satisfy the Lorenz condition:
\begin{equation}\label{eq:Lorenz}
\nabla_\nu (X_m)^\nu = 0 \; .
\end{equation}

In order to set the basic ingredients for future time evolutions, we employ the 3+1 decomposition of the spacetime and the matter fields. The former is done as presented in~\cite{Alcubierre08a}, by foliating the spacetime into space-like hypersurfaces $\Sigma_t$ with 3-metric $\gamma_{ij}$, and the latter by splitting the Proca fields into scalar $\Phi_m$ and 3-vector $(a_m)_i$ potentials, and defining 3-dimensional ``electric'' and ``magnetic'' fields as follows~\cite{Alcubierre:2009ij,Zilhão:2015}:
\begin{alignat}{2}
\label{3+1:def:potentials}
\Phi_m &:= -n^\mu (X_m)_\mu \; , \qquad
& (a_m)_i &:= \tensor{\gamma}{^\mu_i} (X_m)_\mu \; ,
\\
\label{3+1:def:electricomagnetic_fields}
(\mathcal{E}_m)_i &:= -n^\mu\tensor{\gamma}{^\nu_i}(W_m)_{\mu\nu} \; , \qquad
& (\mathcal{B}_m)_i &:= -n^\mu\tensor{\gamma}{^\nu_i}(W^*_m)_{\mu\nu} \; ,
\end{alignat}
where $n^\mu$ is the unit time-like normal vector to the spatial hypersurfaces $\Sigma_t$, $\gamma^\mu_\nu := \delta^\mu_\nu + n^\mu n_\nu$ is the projection operator onto $\Sigma_t$ defined from the 3-metric, and where $(W^*_m)^{\mu\nu} := - E^{\mu\nu\alpha\beta}(W_m)_{\alpha\beta}/2$  denotes the standard dual of the field strength tensor (using the convention that $E^{0123} = -1/\sqrt{-g}$ and $E_{0123} = \sqrt{-g}$).

In this 3+1 formalism, the stress-energy tensor is also decomposed by the following projections: the energy density $\rho:=n^\mu n^\nu T_{\mu\nu}$, the momentum density $j_i := - \tensor{\gamma}{^\mu_i}n^\nu T_{\mu\nu}$, and the spatial stress tensor $S_{ij}:= \tensor{\gamma}{^\mu_i}\tensor{\gamma}{^\nu_j}T_{\mu\nu}$. In terms of the 3+1 decomposition of the Proca fields, these have the form:
\begin{eqnarray}
\rho &=& \sum^N_{m=1}\frac{1}{8\pi}\left\{\ (\mathcal{E}_m)_i (\bar{\mathcal{E}}_m)^i + (\mathcal{B}_m)_i (\bar{\mathcal{B}}_m)^i + \mu^2[\Phi_m \bar{\Phi}_m + (a_m)_i(\bar{a}_m)^i ]\ \right\} \label{exp:3+1:Proca_energy_density}\; ,  \\
j^i &=& \sum^N_{m=1} \frac{1}{8\pi} \left\{\ \tensor{E}{^i_{jk}}(\bar{\mathcal{E}}_m)^j (\mathcal{B}_m)^k + \mu^2 (a_m)^i \bar{\Phi}_m + c.c. \ \right\} \label{exp:3+1:Proca_momentum_density} \; , \\ 
S_{ij} &=& \sum^N_{m=1}\frac{1}{8\pi}\left\{\ \gamma_{ij}[\ (\mathcal{E}_m)_k (\bar{\mathcal{E}}_m)^k + (\mathcal{B}_m)_k (\bar{\mathcal{B}}_m)^k \ ] - [ \ (\mathcal{B}_m)_i (\bar{\mathcal{B}}_m)_j + (\mathcal{E}_m)_i (\bar{\mathcal{E}}_m)_j + c.c. \ ]\right. \nonumber \\  
& & \ \ \ \ \ \ \  + \left. \ \mu^2[\ ( (a_m)_i(\bar{a}_m)_j + c.c. ) - \gamma_{ij}( (a_m)_k(\bar{a}_m)^k - \Phi_m \bar{\Phi}_m)\ ]\ \right\}  \label{exp:3+1:Proca_stress_tensor} \; .
\end{eqnarray}

If we now consider a fixed parameter $\ell \in \mathbb{N}_0$ that plays the role of an angular momentum number, such that $N=2\ell+1$, and consistently re-label the matter fields with $m = -\ell,...,\ell\:$, then a spherically symmetric spacetime is obtained by considering the following ansatz for the potential and electromagnetic fields:
\begin{align}
\label{ansatz:Phi}
    \Phi_m(x) &= \phi_{\ell}(r,t)Y^{\ell m}(\theta,\varphi) \; , \\
\label{ansatz:vector_potential}
    (a_m)_i(x) &= (\ \aleph_\ell(r,t) Y^{\ell m}, \ \beth_\ell(r,t) \partial_\theta Y^{\ell m}, \  \beth_\ell(r,t) \partial_\varphi Y^{\ell m} \ ) \; , \\
\label{ansatz:electric_field}
    (\mathcal{E}_m)_i(x) &= (\ \epsilon_\ell(r,t) Y^{\ell m}, \ \xi_\ell(r,t) \partial_\theta Y^{\ell m}, \ \xi_\ell(r,t) \partial_\varphi Y^{\ell m} \ ) \; , \\
\label{ansatz:magnetic_field}
    (\mathcal{B}_m)^i(x) &=  \left( \ 0, \ \zeta_\ell(r,t) \: \frac{\partial_\varphi Y^{\ell m}}{\sin\theta}, \ - \zeta_\ell(r,t) \: \frac{\partial_\theta Y^{\ell m}}{\sin \theta} \ \right) \; ,
\end{align}
where $Y^{\ell m}(\theta,\varphi)$ denotes the standard spherical harmonics, and the amplitudes $(\phi_\ell, \aleph_\ell, \beth_\ell, \epsilon_\ell, \xi_\ell, \zeta_\ell)$ are the same for all values of $m$. As shown in detail in Lazarte's master's thesis~\cite{Lazarte:2023a}, our ansatz leads to a total stress-energy tensor with spherical symmetry consistent with a spacetime with the same symmetry. This can be shown by replacing \linebreak \eqref{ansatz:Phi}-\eqref{ansatz:magnetic_field} into \eqref{exp:3+1:Proca_energy_density}-\eqref{exp:3+1:Proca_stress_tensor}, and eliminating the angular dependence by using the following identities for the spherical harmonics (which were presented and proved in the appendix of the original paper on $\ell$-boson stars~\cite{Alcubierre:2018}):
\begin{eqnarray}
\sum^\ell_{m=-\ell} |Y^{\ell m}(\theta,\varphi)|^2 &=& 1 \; , \label{prop:SphericalHarmonics:1} \\
\sum^{\ell}_{m=-\ell} \bar{Y}^{\ell m}\partial_\theta Y^{\ell m} = \sum^{\ell}_{m=-\ell} \bar{Y}^{\ell m}\partial_\varphi Y^{\ell m} &=& 0 \; ,  \label{prop:SphericalHarmonics:2} \\
\sum^\ell_{m=-\ell}\left(\partial_\theta Y^{\ell m} \partial_\theta \bar{Y}^{\ell m} + \frac{1}{\sin^2{\theta}}\partial_\varphi Y^{\ell m}\partial_\varphi \bar{Y}^{\ell m}\right) &=& \ell(\ell+1) \; , \label{prop:SphericalHarmonics:3} \\ 
\sum^\ell_{m=-\ell} \partial_\theta Y^{\ell m}\partial_\varphi\bar{Y}^{\ell m} = \sum^\ell_{m=-\ell} \partial_\varphi Y^{\ell m}\partial_\theta \bar{Y}^{\ell m} &=& 0\ . \label{prop:SphericalHarmonics:4.1} \\   
\sum^\ell_{m=-\ell} \partial_\theta Y^{\ell m}\partial_\theta\bar{Y}^{\ell m} = \frac{1}{\sin^2{\theta}}\sum^\ell_{m=-\ell} \partial_\varphi Y^{\ell m}\partial_\varphi\bar{Y}^{\ell m} &=& \frac{\ell(\ell+1)}{2} \: . \label{prop:SphericalHarmonics:4.2}
\end{eqnarray}

 It should be noted that, in comparison with~\cite{Alcubierre:2018}, the spherical harmonics used in this work are rescaled as \linebreak $Y^{\ell m} \rightarrow [4\pi/(2\ell+1)]^{1/2} Y^{\ell m}$.
 
 By choosing to write the spherically symmetric metric in the form (already adapted to the use of the BSSN formulation~\cite{Shibata95,Baumgarte:1998te}):
\begin{equation}\label{spherically:3-metric}
    ds^2 = -\alpha^2(t,r)dt^2 +  \psi^4(t,r)\left(A(t,r)dr^2 + r^2B(t,r)d\Omega^2\right) \; ,
\end{equation}
then the total stress-energy tensor can be expressed by its 3+1 decomposition:
\begin{equation}\label{expr:3+1:stress-energy tensor}
    T_{\mu\nu} = n_\mu n_\nu \rho + n_\mu j_\nu + j_\nu n_\nu + S_{\mu\nu} \; , 
\end{equation}
with the following non-zero contributions:
\begin{eqnarray}
\rho &=& \frac{1}{8\pi} \left\{ \ \frac{1}{\psi^4 A}|\epsilon_\ell|^2  + \mu^2\left[\frac{1}{\psi^4 A}|\aleph_\ell|^2 + |\phi_\ell|^2\right] \right. \nonumber \\
& & \ \ \ \ + \left. \ell(\ell+1) \left[\frac{1}{\psi^4 B r^2} \left( |\xi_\ell|^2 + \mu^2 |\beth_\ell|^2 \right) + \psi^4 B r^2 |\zeta_\ell|^2 \right] \ \right\} \; , \label{expr:spherically:Proca:energy_density} \\
\nonumber \\
j_r &=& \frac{1}{8\pi} \left\{ -\ell(\ell+1)A^{1/2} \psi^2\bar{\xi}_\ell \zeta_\ell + \mu^2\aleph_\ell \bar{\phi}_\ell + c.c. \right\} \; , \label{expr:spherically:Proca:radial_momentum_density} \\ 
\nonumber \\
S_{rr} &=& \frac{\psi^4 A}{8\pi} \left\{ -\frac{1}{\psi^4 A}|\epsilon_\ell|^2  + \mu^2\left[\frac{1}{\psi^4 A}|\aleph_\ell|^2 + |\phi_\ell|^2\right] \right. \nonumber \\
& & \ \ \ \ + \left. \ell(\ell+1)\left[\frac{1}{\psi^4 B r^2}\left(|\xi_\ell|^2 - \mu^2 |\beth_\ell|^2\right) + \psi^4 B r^2 |\zeta_\ell|^2 \right] \ \right\} \; , \label{expr:spherically:Proca:radial_stress_tensor} \\
\nonumber \\
S_{\theta \theta} &=& \frac{\psi^4 B r^2}{8\pi} \left\{ \ \frac{1}{\psi^4 A} |\epsilon_\ell|^2  - \mu^2\left[ \frac{1}{\psi^4 A}| \aleph_\ell|^2 - |\phi_\ell|^2 \right] \right\} \label{expr:spherically:Proca:polar_stress_tensor} \; , \\
\nonumber \\
S_{\varphi\varphi} &=& \frac{\psi^4 B r^2 \sin^2{\theta}}{8\pi} \left\{ \ \frac{1}{\psi^4 A}|\epsilon_\ell|^2  - \mu^2\left[\frac{1}{\psi^4 A}|\aleph_\ell|^2 - |\phi_\ell|^2\right] \right\}  \label{expr:spherically:Proca:azimutal_stress_tensor} \; .
\end{eqnarray}

Although the spacetime is spherically symmetric, we must stress the fact that for non-trivial values of $\ell$ there are non-zero ``magnetic'' field contributions to the stress-energy tensor coming from the $\zeta_\ell(r,t)$ terms, in contrast with the one-field spherically symmetric configuration case, which in our proposed ansatz~\eqref{ansatz:Phi}-\eqref{ansatz:magnetic_field} implies \linebreak $\ell=0: Y^{00}=1,\   \partial_\theta Y^{00} =\partial_\varphi Y^{00}=0\ \Rightarrow\  (\mathcal{B}_0)_i = (0,0,0)$. In particular, this case includes the spherical Proca stars where the magnetic field vanishes~\cite{BRITO2016291}. Again, for non-trivial values of $\ell$, the magnetic fields in our proposal would seem to be singular at $\theta=0,\pi$ due to the presence of a division by a sine function in the ansatz~\eqref{ansatz:magnetic_field}.  However, as shown in the first appendix of \cite{Lazarte:2023a}, such a division only introduces coordinate singularities in the azimuthal components of the 3-vector magnetic field with $m=\pm 1$, that can be removed by changing to Cartesian coordinates.

From the definition of $(\mathcal{B}_m)_i$ in~\eqref{3+1:def:electricomagnetic_fields} one can show that $(\mathcal{B}_m)^i = E^{ijk}\partial_j (a_m)_k $, which by using our ansatz leads to:
\begin{equation}
\zeta_\ell = \frac{( \ \aleph_\ell - \partial_r \beth_\ell\ )}{A^{1/2} B \psi^6 r^2} \; .
\end{equation}
This last property allows us to express the evolution of the Proca fields in terms of the dynamical variables \linebreak $\{ \Phi_m, (a_m)_i, (\mathcal{E}_m)^i \}$, which by using our ansatz can be replaced with $\{\phi_\ell, \aleph_\ell, \beth_\ell, \epsilon_\ell, \xi_\ell\}$.  Considering that the Lorenz condition~\eqref{eq:Lorenz} prescribes the evolution of $\Phi_m$, the definition of the $W_{\mu\nu}$ tensor prescribes the evolution of $(a_m)_i$, and the spatial projection of the equation of motion (\ref{eq:field:Proca}) yields the evolution equation of $(\mathcal{E}_m)^i$, one can find the following system of evolution equations:
\begin{eqnarray}
\partial_t \phi_\ell &=& - \frac{\alpha}{A\psi^4} \left[ (\partial_r\aleph_\ell) + \aleph_\ell \left(\frac{2}{r} - \frac{\partial_r A}{2A} + \frac{\partial_r B}{B} + \frac{2\partial_r \psi}{\psi} + \frac{\partial_r \alpha}{\alpha}\right) \right] + \alpha \: \frac{\ell (\ell+1)}{B \psi^4 r^2} \: \beth_\ell + \alpha K \phi_\ell \; , \label{eq:evol:radial:phi} \\
\partial_t \aleph_\ell &=& - \alpha \epsilon_\ell - \partial_r (\alpha \phi_\ell) \; , \label{eq:evol:radial:aleph} \\
\partial_t \beth_\ell &=& - \alpha \xi_\ell  - \alpha \phi_\ell \; , \label{eq:evol:radial:beth} \\
\partial_t \epsilon_\ell &=& \alpha \: \frac{\ell(\ell+1)}{B  \psi^4 r^2} \left( \aleph_\ell - \partial_r \beth_\ell \right)   + \alpha \left[ \mu^2\aleph_\ell + \left( K - K^r_r \right) \epsilon_\ell \right] \; , \label{eq:evol:radial:epsilon} \\
\partial_t \xi &=& \frac{\alpha}{A\psi^4} \left[ \partial_r \left( \aleph_\ell
- \partial_r \beth_\ell \right) + \left( \aleph_\ell -\partial_r \beth_\ell \right)
\left( \frac{\partial_r \alpha}{\alpha} - \frac{\partial_r A}{2A}
- \frac{2\partial_r \psi}{\psi} \right) \right] + \alpha \left[ \mu^2 \beth_\ell
+ \left( K - K^\theta_\theta \right) \xi_\ell \right] \; . \label{eq:evol:radial:xi}
\end{eqnarray}
In the above equations $K_{ij}$ are the components of the extrinsic curvature tensor of the spatial hypersurfaces, with $K$ its trace. Furthermore, the normal projection of the equation of motion~\eqref{eq:field:Proca} yields a ``Gauss'' constraint equation of the form:
\begin{equation}\label{eq:constr:Gauss:electric:radial}
0 = \frac{1}{A\psi^4}\left[ \partial_r \epsilon_\ell + \epsilon_\ell\left(\frac{2}{r}
- \frac{\partial_r A}{2A} + \frac{\partial_r B}{B} + \frac{2 \partial_r \psi}{\psi} \right) \right] - \frac{\ell(\ell+1)}{B \psi^4 r^2} \: \xi_\ell + \mu^2 \phi_\ell \; .
\end{equation}

As can be easily seen, the resulting system of evolution and constraint equations only has spatial derivatives in the radial coordinate. The angular derivatives are not present because they combine to form the usual Laplacian operator in angular coordinates applied to $Y^{\ell m}$, which can be replaced using the following identity:
\begin{equation}
\label{prop:SphericalHarmonics:Laplacian}
\frac{1}{\sin{\theta}} \: \partial_\theta (\sin{\theta} \: \partial_\theta Y^{\ell m}) + \frac{1}{\sin^2{\theta}} \: \partial^2_\varphi Y^{\ell m} = - \ell(\ell+1)Y^{\ell m} \; ,
\end{equation}
thus, generating the centrifugal terms $\ell(\ell+1)/r^2$ present in the equations~\eqref{eq:evol:radial:phi}-\eqref{eq:constr:Gauss:electric:radial}. 

As a final comment, it is well known that the Proca evolution equations (and indeed the standard Maxwell evolution equations) possess non-propagating constraint violating modes~\cite{Zilhão:2015}, which may lead to instabilities in a numerical simulation~\cite{Alcubierre99e}. In order to fix this, modifications have been proposed as the analogous version of the BSSN formulation for electromagnetism~\cite{Knapp:2002fm}. However, these constraint violating modes do not seem to cause any problem in the spherically symmetric case, so we will not consider them here.


\section{$\ell$-Proca stars}
\label{sec:procastar}

As we are looking for stationary bosonic stars solutions which would correspond to the symmetry-enhanced solutions in a family of multi-field and multi-frequency Proca stars, we impose in the former ansatz \eqref{ansatz:Phi}-\eqref{ansatz:electric_field} the same harmonic time-dependency for the $2\ell+1$ Proca fields by employing the following prescription:
\begin{equation}
\label{ansatz:static:spherically:Proca_potencials}
\phi_\ell(r,t) = \varphi_\ell(r) e^{-i\omega t} \; , \quad
\aleph_\ell (r,t) = i a_\ell(r) e^{-i\omega t} \; , \quad
\beth_\ell (r,t) = i b_\ell(r) e^{-i\omega t} \; , 
\end{equation}
\begin{equation}
\label{ansatz:static:spherically:Proca_fields}
\epsilon_\ell(r,t) = e_\ell(r) e^{-i\omega t} \; , \quad
\xi_\ell(r,t) = d_\ell(r)  e^{-i\omega t} \; ,
\end{equation}
where $\omega$ is a real frequency parameter, and with the radial profiles $\varphi_\ell(r), a_\ell(r),  b_\ell(r), e_\ell(r),$ $d_\ell(r)$ also real-valued. Since this ansatz leads to a time-independent stress-energy tensor, these bosonic stars result in static spacetime solutions of the spherically symmetric Einstein-(multi)Proca system. Furthermore, by replacing our ansatz in the evolution and constraint equations, and considering that for a static spacetime the extrinsic curvature tensor vanishes, one obtains the following self-consistent system of purely radial equations:
\begin{align}
\omega \varphi_\ell &= \frac{1}{A^{1/2}B\psi^6 r^2} \left( \alpha \: \frac{\psi^2 B r^2}{A^{1/2}} \: a_\ell \right)' - \alpha \: \frac{\ell(\ell+1)}{B\psi^4 r^2} \: b_\ell \; , \label{eq:initial_data:first_order:phi} \\
\omega a_\ell &= - \alpha e_\ell - (\alpha \varphi_\ell)' \; , \label{eq:initial_data:first_order:a} \\
\omega b_\ell &= - \alpha d_\ell - \alpha \varphi_\ell\; , \label{eq:initial_data:first_order:b} \\
\omega e_\ell &= - \alpha \: \frac{\ell(\ell+1)}{B\psi^4 r^2}(a_\ell-b'_\ell) - \alpha \mu^2 a_\ell \; , \label{eq:initial_data:first_order:e} \\
\omega d_\ell &= - \frac{1}{A^{1/2}\psi^2} \left(\alpha\frac{(a_\ell-b'_\ell)}{A^{1/2}\psi^2} \right)' - \alpha \mu^2 b_\ell \; , \label{eq:initial_data:first_otder:d} \\
0 &= \frac{1}{A^{1/2}B\psi^6r^2} \left( \frac{\psi^2Br^2}{A^{1/2}} e_\ell \right)' - \frac{\ell(\ell+1)}{B\psi^4 r^2} \: d_\ell + \mu^2 \varphi_\ell \; , \label{eq:initial_data:Gauss_constraint}
\end{align}
where here the prime denotes the radial derivative. By ``self-consistency'' here we mean the fact that the 5 radial equations~\eqref{eq:initial_data:first_order:phi}-\eqref{eq:initial_data:first_otder:d} that emerged from the evolution equations ensure that the Gauss constraint~\eqref{eq:initial_data:Gauss_constraint} is automatically satisfied. This can be easily proved by using equations~\eqref{eq:initial_data:first_order:e}-\eqref{eq:initial_data:first_otder:d}, that result from the evolution equations for the 3-dimensional ``electric'' field \eqref{eq:evol:radial:epsilon}-\eqref{eq:evol:radial:xi}, to show that the Lorenz condition~\eqref{eq:initial_data:first_order:phi} reduces to the ``Gauss'' constraint~\eqref{eq:initial_data:Gauss_constraint}. Although this feature seems trivial, it was crucial in order to propose the angular dependence of the vector components in our ansatz~\eqref{ansatz:Phi}-\eqref{ansatz:magnetic_field}. In fact, we initially tried somewhat simpler versions for our ansatz that were ruled out because, even though they generated a spherically symmetric stress--energy tensor, they resulted in evolution equations that did not satisfy the Gauss constraint once we imposed the harmonic time dependence. In this sense, our proposed ansatz is self-consistent since, if one can find solutions for the radial equations~\eqref{eq:initial_data:first_order:phi}-\eqref{eq:initial_data:first_otder:d}, they are guaranteed to satisfy the Gauss constraint~\eqref{eq:initial_data:Gauss_constraint}.

Notice that the evolution equations~\eqref{eq:initial_data:first_order:a}-\eqref{eq:initial_data:first_order:b}  give us a way to derive the amplitudes $(e_\ell, d_\ell)$ directly from $(F_\ell,
a_\ell, b_\ell)$, where we have defined $F_\ell:=\alpha \varphi_\ell$. Indeed, one finds:
\begin{equation}\label{expr:simplication:e-d}
e_\ell = -\frac{1}{\alpha} \left( F_\ell' + \omega a_\ell \right) \; , \qquad
d_\ell = -\frac{1}{\alpha} \left( F_\ell  + \omega b_\ell \right) \; .
\end{equation}
As a consequence, in order to obtain our solutions we first have to compute the radial amplitudes of the potential fields $(F_\ell,a_\ell,b_\ell)$. If we now look at the potential one form:
\begin{equation}
X_m = e^{-i \omega t} \left[ -F_\ell(r) Y^{\ell m} dt + i a_\ell(r) Y^{\ell m} dr
+ i b_\ell(r) \left( \partial_\theta Y^{\ell m} d \theta 
+ \partial_\varphi Y^{\ell m} d \varphi \right) \right] \; ,
\label{ansatz:potential:1-form}
\end{equation}
it is easy to see that our ansatz introduces a dependency on the angular momentum number $\ell$ such that in the particular case of $\ell=m=0$ it reduces to the ansatz for a standard Proca star with zero angular momentum, with Eqs.~\eqref{eq:initial_data:first_order:b} and \eqref{eq:initial_data:first_otder:d} no longer being part of the system due to the fact that the angular components of the 3-vector fields in~\eqref{ansatz:vector_potential}-\eqref{ansatz:magnetic_field} vanish.  Equations~\eqref{eq:initial_data:Gauss_constraint} and \eqref{eq:initial_data:first_order:a}, with $e_\ell$ given by \eqref{expr:simplication:e-d} and \eqref{eq:initial_data:first_order:e} respectively, then reduce to the usual equations for a standard Proca star (notice that in the notation of the original paper on Proca stars~\cite{BRITO2016291} we have $f(r):= -F_{\ell=0}(r)$ and $g(r):= a_{\ell=0}(r)$, with the spacetime metric written in polar-areal coordinates such that $\psi^4 B=1$, $\psi^4 A=1/N$ and $\alpha^2=\sigma^2 N$).  Thus, our proposal represents a generalization of Proca stars to the case of a non-zero total angular momentum parameterized by $\ell$. For this reason, in analogy with the scalar case~\cite{Alcubierre:2018}, we refer to these objects as $\ell$-Proca stars. \\

As discussed below, we will solve numerically the system of radial equations for the $\ell$-Proca stars using spectral methods, for which we need to express those equations as a system of ordinary differential equations (ODE´s) of second order. In order to do this we first assume that our spatial metric is given in isotropic coordinates such that $A=B=1$ in the line element~\eqref{spherically:3-metric} above, so that it reduces to:
\begin{equation}\label{spherically:static:3-metric}
ds^2 = -\alpha^2(r)dt^2 + \psi^4(r) \left( dr^2 + r^2 d\Omega^2 \right) \; .
\end{equation}
As this spacetime metric is static, the extrinsic curvature $K_{ij}$ vanishes.  This implies that: (i) the Hamiltonian constraint results in a second order ODE for the conformal factor $\psi$, and (ii) $K = \partial_t K = 0$, which is equivalent to asking for the maximal slicing condition that leads to a second order ODE for the lapse function $\alpha$.  Also, by making use of our ansatz~\eqref{ansatz:Phi}-\eqref{ansatz:magnetic_field} with the radial and time dependency specified by~\eqref{ansatz:static:spherically:Proca_potencials}-\eqref{ansatz:static:spherically:Proca_fields}, we can substitute the potential one-form~\eqref{ansatz:potential:1-form} into the Proca equations~\eqref{eq:field:Proca} written in the following form:
\begin{equation}\label{eq:Proca_field:second_order}
    \nabla_\mu\nabla^\mu (X_m)_\nu - R_{\mu\nu}(X_m)^\mu - \mu^2 (X_m)_\nu = 0 \; ,
\end{equation}
and obtain from its components  $\nu = t, r, \theta  \ (\text{or}\ \varphi)$ second order ODE´s for the radial amplitudes $(F_\ell, a_\ell, b_\ell)$, respectively. The resulting system of five second order ODE´s has the final form:
\begin{align}
0 &=  F''_\ell + \frac{2 F'_\ell}{r} - \frac{\ell(\ell+1)}{r^2} \: F_\ell 
+ \frac{2\psi'}{\psi} \: F'_\ell - \frac{\alpha'}{\alpha} \left( F'_\ell+2\omega a_\ell \right) - \psi^4 \left( \mu^2 - \frac{\omega^2}{\alpha^2} \right) F_\ell \; , \label{eq:initial_data:second_order:F} \\
0 &= a''_\ell + \frac{2a'_\ell}{r} - \frac{2a_\ell}{r^2} - \frac{\ell(\ell+1)}{r^2} \: a_\ell
+ \frac{2\ell(\ell+1)}{r^3} \: b_\ell - \frac{2a_\ell}{r} \left( \frac{\alpha'}{\alpha}
+ \frac{6\psi'}{\psi} \right) + \frac{4\ell(\ell+1)}{r^2} \: \frac{\psi'}{\psi} \: b_\ell
+ \left(\frac{\alpha'}{\alpha} - \frac{2\psi'}{\psi} \right )a'_\ell \nonumber \\
& +\frac{2\psi^4}{\alpha^2} \: \frac{\alpha'}{\alpha} \: \omega F_\ell
- a_\ell \left[ 10 \left( \frac{\psi'}{\psi} \right)^2 + \frac{6\alpha'\psi'}{\alpha\psi}
+ \left(\frac{\alpha'}{\alpha} \right)^2 \right] - \psi^4 \left( \mu^2 - \frac{\omega^2}{\alpha^2} \right) a_\ell + 4\pi\psi^4 a_\ell S \; , \label{eq:initial_data:second_order:a}\\ 
0 &= b''_\ell - \frac{\ell(\ell+1)}{r^2} \: b_\ell + \frac{2 a_\ell}{r} + \left( \frac{\alpha'}{\alpha}-\frac{2\psi'}{\psi} \right) b'_\ell + \frac{4\psi'}{\psi} \: a_\ell
- \psi^4 \left( \mu^2 - \frac{\omega^2}{\alpha^2} \right) b_\ell \; , \label{eq:initial_data:second_order:b}\\
0 &= \psi'' + \frac{2\psi'}{r} + 2 \pi \psi^5 \rho \; , \label{eq:initial_data:second_order:psi}\\
0 &= \alpha'' + \frac{2\alpha'}{r} +\frac{2\alpha'\psi'}{\psi}
- 4 \pi \alpha \psi^4 \left( S + \rho \right) \; . \label{eq:initial_data:second_order:lapse}
\end{align}
where the energy density $\rho$ and the trace of the spatial stress tensor $S=\tensor{S}{^i_i}$ are computed from the matter source terms~\eqref{expr:spherically:Proca:energy_density}-\eqref{expr:spherically:Proca:azimutal_stress_tensor} using our ansatz for the $\ell$-Proca star and the spacetime metric~\eqref{spherically:static:3-metric}: 
\begin{align}
\rho &= \frac{1}{8\pi} \left\{ \ \frac{(F'_\ell+\omega a_\ell)^2}{\alpha^2\psi^4} + \mu^2 \left[ \frac{F_\ell^2}{\alpha^2} + \frac{a_\ell^2}{\psi^4} \right] + \frac{\ell(\ell+1)}{\psi^4 r^2} \left[ \frac{(F_\ell + \omega b_\ell)^2}{\alpha^2} + \mu^2 b_\ell^2 + \frac{(a_\ell - b'_\ell)^2}{\psi^4} \right] \right\} \label{ell_ProcaStar:energy_density}\; ,\\
S & = \frac{1}{8\pi} \left\{ \frac{(F'_\ell+\omega a_\ell)^2}{\alpha^2\psi^4} + \mu^2 \left[ \frac{3F_\ell^2}{\alpha^2} -\frac{a_\ell^2}{\psi^4} \right] + \frac{\ell(\ell+1)}{\psi^4  r^2} \left[ \frac{(F_\ell + \omega b_\ell)^2}{\alpha^2} - \mu^2 b_\ell^2 + \frac{(a_\ell - b'_\ell)^2}{\psi^4 } \right] \right\}\; . \label{ell_ProcaStar:trace:stress_tensor} 
\end{align}
As mentioned above, the second order derivatives in the angular coordinates coming from the 3-dimensional Laplacian operator are replaced by centrifugal terms using~\eqref{prop:SphericalHarmonics:Laplacian}. Imposing appropriate boundary conditions at infinity (see below), this system of equations forms a nonlinear eigenvalue problem for the frequency $\omega$.


\section{Local solutions close the origin}
\label{sec:localsolutions}

An analysis of the Proca equations at the boundaries of the radial domain is required in order to impose correctly the boundary conditions for the matter fields. For the outer boundary conditions, it is easy to derive from the Eqs.~\eqref{eq:initial_data:second_order:F}-\eqref{eq:initial_data:second_order:b} that the matter fields have exponential solutions far away as long as we have $\omega < \mu$.  The physical solutions should correspond to those that decay for large values of $r$, as is expected for a compact object. This is in fact what gives rise to the eigenvalue problem, as arbitrary values of the frequency $\omega$ would typically correspond to exponentially growing solutions.

On the other hand, in order to impose regularity conditions at the origin we perform an analysis for small radius as follows: It is a general result in spherical symmetry that every regular function must have a well-defined parity, therefore from the radial equations for the metric components~\eqref{eq:initial_data:second_order:psi}-\eqref{eq:initial_data:second_order:lapse} one can easily show that the metric functions $(\alpha,\psi)$ must be even. In terms of power series we must have for small $r$:
\begin{align}
\label{alpha:local_solution}
\alpha &= \alpha_0 + \alpha_2 r^2 + \mathcal{O}(r^4) \; , \\
\label{psi:local_solution}
\psi &= \psi_0 + \psi_2 r^2 + \mathcal{O}(r^4) \; .
\end{align}
with $(\psi_0,\alpha_0)$ constants, and where the coefficients $(\psi_2,\alpha_2)$ can be obtained by replacing~\eqref{psi:local_solution}-\eqref{alpha:local_solution} in the radial equations~\eqref{eq:initial_data:second_order:psi}-\eqref{eq:initial_data:second_order:lapse}, resulting in the following conditions:
\begin{equation}\label{expr:second_order_approx}
\frac{\psi_2}{\psi_0} = -\frac{\pi}{3} \: \psi^4_0 \rho_0\; , \qquad
\frac{\alpha_2}{\alpha_0} = \frac{2\pi}{3} \: \psi^4_0 \left( S_0 + \rho_0 \right) \; ,
\end{equation}
with $\rho_0 := \rho(r=0)$ and $S_0:=S(r=0)$.

We can now obtain the radial equations for the matter variables in the vicinity of the origin by substituting the expansions~\eqref{psi:local_solution}-\eqref{alpha:local_solution} in the system of equations for the matter fields~\eqref{eq:initial_data:second_order:F}-\eqref{eq:initial_data:second_order:b}, resulting in the following approximated form for small radius:
\begin{align}
F''_\ell + \frac{2 F'_\ell}{r} + \left[ \psi^4_0 \left( \frac{\omega^2}{\alpha^2_0}
- \mu^2 \right) - \frac{\ell(\ell+1)}{r^2}\right] F_\ell
&\approx 0 \: , \\
a''_\ell + \frac{2a'_\ell}{r} + \left[ \psi^4_0 \left( \frac{\omega^2}{\alpha^2_0}
- \mu^2 \right) - 4 \left( \frac{\alpha_2}{\alpha_0} + \frac{6\psi_2}{\psi_0} \right)
+ 4 \pi \psi^4_0 S_0  - \frac{(2+\ell(\ell+1))}{r^2} \right] a_\ell
&\approx - 2 \left[ \frac{1}{r^2} + 4 \left( \frac{\psi_2}{\psi_0} \right) \right] \frac{\ell(\ell+1)}{r} \: b_\ell \: , \label{eq:local:a_aux} \\
b''_\ell + \left[ \psi^4_0\left(\frac{\omega^2}{\alpha^2_0}-\mu^2\right) - \frac{\ell(\ell+1)}{r^2}\right] b_\ell &\approx - \frac{2 a_\ell}{r} \: . 
\end{align}
In order to simplify the equation for $a''_\ell$ above we use equation~\eqref{eq:initial_data:first_order:phi} adapted to the metric given by~\eqref{spherically:static:3-metric} to express $b_\ell$ in the right hand side of this equation in terms of $F_\ell$. Then, by taking into account the expressions for the second order coefficients of $\alpha$ and $\psi$, Eq.~\eqref{expr:second_order_approx}, the system can be shown to reduce to:
\begin{align}
F''_\ell + \frac{2 F'_\ell}{r} + \left[ \psi^4_0 \left( \frac{\omega^2}{\alpha^2_0}-\mu^2\right) 
- \frac{\ell(\ell+1)}{r^2} \right]F_\ell &\approx 0 \; ,
\label{eq:initial_data:second_order:local_solution:F} \\
a''_\ell + \frac{4a'_\ell}{r} + \left[ \psi^4_0 \left(\frac{\omega^2}{\alpha^2_0} - \mu^2 + \frac{4\pi}{3}S_0 \right) + \frac{(2-\ell(\ell+1))}{r^2} \right] a_\ell &\approx
\frac{2 \psi^4_0 \omega}{\alpha^2_0} \: \frac{F_\ell}{r} \: ,
\label{eq:initial_data:second_order:local_solution:a} \\ 
b''_\ell + \left[ \psi^4_0 \left( \frac{\omega^2}{\alpha^2_0} - \mu^2 \right)
- \frac{\ell(\ell+1)}{r^2} \right] b_\ell &\approx - \frac{2 a_\ell}{r} \; . \label{eq:initial_data:second_order:local_solution:b}
\end{align}

Homogeneous solutions of this system can be obtained using Bessel functions, and particular solutions using the method of variation of parameters. By imposing regularity at the origin, the general solution of this system takes the form:
\begin{align}
F_\ell & = c_1 r^\ell - 2 \left( \frac{\kappa_1}{2} \right)^2 c_1 \: \frac{r^{\ell+2}}{(2\ell+3)} + O(r^{\ell+4}) \; , \label{F:local_solution} \\
a_\ell & = \ell c_2 r^{\ell - 1} + \left[ \frac{\psi^4_0 \omega}{\alpha^2_0} \: c_1
- 2\ell \left( \frac{\kappa_2}{2}\right)^2 c_2 \right] \frac{r^{\ell +1 }}{(2\ell+3)} + O(r^{\ell+3}) \; , \label{a:local_solution} \\
b_\ell & = c_2 r^\ell - \left[ \frac{\psi^4_0\omega}{\alpha_0^2}c_1 + 2 (2\ell+3) \left( \frac{\kappa_1}{2} \right)^2 c_2 - 2 \ell \left( \frac{\kappa_2}{2 }\right)^2 c_2 \right] \frac{r^{\ell+2}}{(\ell +1 )(2\ell + 3)} + O(r^{\ell+4}) \; , \label{b:local_solution}
\end{align}
with $(c_1,c_2)$ arbitrary parameters, and where $(\kappa_1,\kappa_2)$ are positive quantities defined as: 
\begin{equation}\label{expr:kappas}
\kappa^2_1 = \psi^4_0 \left( \frac{\omega^2}{\alpha^2_0} - \mu^2 \right) \; , \qquad
\kappa^2_2 = \psi^4_ 0\left( \frac{\omega^2}{\alpha^2_0} - \mu^2 + \frac{4\pi}{3} \: S_0 \right) \; . 
\end{equation}
The solutions~\eqref{F:local_solution}-\eqref{b:local_solution} give us the local behavior of the matter functions near the origin. These solutions reduce to the standard case of Proca stars~\cite{BRITO2016291} for $\ell=0$, as follows:
\begin{equation}\label{local:ell=0}
F_{\ell=0} = c_1 + \frac{c_1}{6}\left(\mu^2-\frac{\omega^2}{\alpha^2_0}\right) r^2
+ \mathcal{O}(r^4) \; , \qquad
a_{\ell=0} = c_1 \left( \frac{\psi_0^4 \omega }{3\alpha^2_0} \right) r
+ \mathcal{O}(r^3) \; .
\end{equation}
Notice that in this case there in no dependence on $c_2$, and the local solutions depend on a single free parameter $c_1$ (in the notation of the original paper on Proca stars~\cite{BRITO2016291} we have $f(r):= -F_{\ell=0}(r)$ and $g(r):= a_{\ell=0}(r)$, with $c_1=-f_0$,  $\psi^4_0=1$ and $\alpha^2_0=\sigma^2_0$.)  However, for non-trivial values of $\ell$ we stress the fact that we must choose two independent parameters $c_1$ and $c_2$ in order to build the general solution, unlike the $\ell=0$ case.

One can compare the above expressions with the local solutions near the origin for $\ell$-boson stars configurations~\cite{Alcubierre:2018} where the radial dependence of the scalar fields has a dominant power of $r^\ell$, thus avoiding singularities that could come from the centrifugal terms in the stress-energy tensor. However, as one can notice above, in the case of the $\ell$-Proca stars the function $a_\ell$ has instead a dominant power of $r^{\ell-1}$.  One could then worry that the centrifugal terms in the stress--energy tensor might not be regular.  However, as one can see from equations~\eqref{ell_ProcaStar:energy_density} and~\eqref{ell_ProcaStar:trace:stress_tensor}, $a_\ell$ only enters into the centrifugal terms through the combination $(a_\ell - b'_\ell)$, and the expansions for small $r$ above imply that:
\begin{equation}
\frac{a_\ell - b'_\ell}{r^2} = \frac{1}{r^2}
\left[ \left( \ell c_2 r^{\ell -1} + \mathcal{O}(r^{\ell+1}) \right)
- \left( \ell c_2 r^{\ell -1} + \mathcal{O}(r^{\ell+1} \right) \right]
= \mathcal{O}(r^{\ell-1}) \; ,
\end{equation}
which is regular for all $\ell \geq 1$ (for $\ell=0$ there is also no problem as the centrifugal term is not present).

On the other hand, the dominant power of $r^{\ell-1}$ in the expansion for $a_\ell$ does mean that the particular case $\ell=1$ deserves some special attention. Equation~\eqref{a:local_solution} implies that for $\ell=1$ the function $a_{\ell=1}$ now takes a constant value at the origin, namely $a_{\ell=1}(r=0)=c_2$.  This is troublesome since in our ansatz the coefficient $a_\ell$ corresponds to the radial component of the 3-vector potential, and as such it should vanish at the origin in order to maintain regularity (a non-zero radial component at the origin implies that the vector field is not differentiable there). In order to avoid this problem one could argue that we should simply take $c_2=0$. Indeed, this was the first thing we tried, and we explored the possibility of solving the $\ell$-Proca star equations for $\ell=1,2,3$ imposing a zero value of $c_2$ on the local matter solutions near the origin. This was done by implementing a routine based on a 1-dimensional shooting method, taking $c_1$ as a free parameter and the frequency $\omega$ as the eigenvalue to be found. However, doing this we were unable to find exponentially decaying solutions for all of the three matter profile functions $(F_\ell,a_\ell,b_\ell)$ simultaneously for any value of the frequency $\omega$.  This suggests that in order to find exponentially decaying solutions far away for all three functions $(F_\ell,a_\ell,b_\ell)$ one must have a non-zero value of $c_2$. This has the consequence that for the particular case of $\ell=1$, our solutions, while still retaining spherical symmetry and remaining finite everywhere, are in fact not regular at the origin and are thus not physical (though one could argue that they correspond to solutions with a point-like source for the Proca field at the origin).  Nevertheless, when we discuss our numerical results below we will also present an example of a solution with $\ell=1$ in order to show that, apart from the fact that $a_{\ell=1}$ is non-zero (but finite) at the origin, such solutions are perfectly well behaved.


\section{Numerical results}
\label{sec:results}


\subsection{Multi-domain Collocation Spectral Method}

As mentioned above, in order to find solutions for our $\ell$-Proca stars one needs to consider two independent parameters at the origin, namely $c_1$ and $c_2$.  This implies that a simple one-dimensional shooting method, like the one used in the case of $\ell$-boson stars in~\cite{Alcubierre:2018}, is no longer adequate to find our solutions.  We have therefore decided to implement a spectral method in order to solve the $\ell$-Proca system of   equations~\eqref{eq:initial_data:second_order:F}-\eqref{eq:initial_data:second_order:lapse}.

The code written for this purpose implements a multi-domain collocation spectral method based on specific details presented in~\cite{grandclement2009:spectralreview,Grandclement2014:rotatingBSs,grandclement2010:kadath,Alcubierre:2022}. The 1-dimensional physical radial space $[0,\infty)$ was split into a finite set of domains. For each radial domain we set a mapping between the radial coordinate $r$ and a numerical coordinate $x\in [-1,1]$, and approximate the radial functions $(F_\ell,a_\ell,b_\ell,\alpha,\psi)$ to their spectral projections onto a finite basis of Chebyshev polynomials $\{T_k(x), k=1,...,N\}$. Then, in each numerical domain, we use  a Gauss--Lobato quadrature and implement the collocation method~\cite{grandclement2009:spectralreview}, which is based on the projection of the system of equations using Lagrange cardinal polynomials defined on the collocation points. The system of nonlinear equations resulting from each numerical domain has as unknowns the coefficients of the spectral projections performed in it.  To find the solutions in the whole space we joined the systems of equations for every domain using matching conditions at interior boundaries. Thus, we obtained one large system of nonlinear equations, which is finally solved by means of a Newton--Raphson iteration method (details of the numerical code can be found in Lazarte's master's thesis~\cite{Lazarte:2023a}).

This multi-domain technique allows us to choose a different mapping between the radial coordinate $r$ and the numerical coordinate $x$ for each domain according to our purposes. In particular, we chose a mapping that compactifies the outer numerical domain in order to include spatial infinity in our numerical grid~\cite{grandclement2009:spectralreview,Grandclement2014:rotatingBSs}, where we impose the following outer boundary conditions: 
\begin{equation}\label{boundary_conditions}
F_\ell(r\rightarrow \infty)=0 \; , \quad
a_\ell(r\rightarrow \infty)=0 \; , \quad
b_\ell(r\rightarrow \infty)=0 \; , \quad 
\alpha(r\rightarrow \infty)=1 \; , \quad
\psi(r\rightarrow \infty)=1 \;  . 
\end{equation}
Notice that in the case of the matter functions $(F_\ell,a_\ell,b_\ell)$, we are not imposing explicitly an exponential decay when $r$ tends to infinity, but rather the actual asymptotic values at infinity itself.  However, as we can verify for our numerical results (see next subsection), the radial equations themselves generate this type of exponential decay. 

To impose the regularity conditions at $r=0$ we restricted the parity of the Chebyshev polynomials defined on the numerical domain that encompasses the origin (a ``nucleus-type'' domain, see~\cite{Grandclement2014:rotatingBSs}). Such a parity restriction depends on the parity of the local solutions close to the origin as determined by the dominant power in their power series expansions.  In particular, for the metric functions we have from~\eqref{psi:local_solution}-\eqref{alpha:local_solution}:
\begin{equation}
\alpha = \alpha_0 + \mathcal{O}(r^2) \; , \qquad
\psi = \psi_0 + \mathcal{O}(r^2) \; .
\end{equation}
We therefore use an even basis for the spectral projection of $(\alpha,\psi)$ in the nucleus-type domain regardless of the value of $\ell$. On the other hand, by taking into account the local solutions for the matter functions~\eqref{F:local_solution}-\eqref{b:local_solution} we have the following dependency for $\ell=0$ (remember  that in this case the function $b_\ell$ does not play a role):
\begin{equation}\label{matter:local:ell=0}
F_\ell \propto 1 \; ,
\qquad a_\ell \propto r \; ,
\end{equation}
while for $\ell>0$ we have instead:
\begin{equation}\label{matter:local:ell>0}
F_\ell \propto r^{\ell} \; ,
\qquad a_\ell \propto r^{\ell-1} \; ,
\qquad b_\ell \propto r^{\ell} \; .
\end{equation}
We then see that the parity of their respective Chebyshev polynomial basis for each of the $(F_\ell, a_\ell, b_\ell)$ will be either even or odd depending the parity of the respective power of $r$.

Our spectral code considered four values for the angular momentum number, namely $\ell=0,1,2,3$. For the particular cases $\ell=2,3\:$, it was necessary to impose additional regularity conditions, as simply restricting the parity of the spectral basis is not enough to ensure that, for functions behaving as $r^n$ with $n>1$, both the function and its first derivative vanish at the origin~\cite{grandclement2010:kadath}. For example, we verified that by only imposing an even spectral basis one finds that either $b_{\ell=2}(0) \neq 0$ or $a_{\ell=3}(0)\neq 0$, due to a non-zero contribution coming from the term $T_0(0)=1$ in the spectral projection.  Similarly, by only imposing an odd spectral basis we obtained $F'_{\ell=3}(0) \neq 0$ or $b'_{\ell=3}(0)\neq 0$, due to a non-zero contribution from $T_1(x)=x$ that implies $T'_1(0)=T_0(0)=1$.  In consequence, in order to guarantee that those functions are zero at the origin we followed the Galerkin technique implemented in~\cite{grandclement2010:kadath}, which consists in choosing a spectral basis whose elements already satisfy the additional constraints one wishes to enforce in the solutions. Thus, for the 2-Proca and 3-Proca stars, we use for  $b_2$ and $a_3$ the even Galerkin basis with elements $G_k(x) = T_{2k+2}(x) -(-1)^{k+1}T_0(x)$, and for $F_3$ and $b_3$ the odd Galerkin basis with elements $G_k(x) = T_{2k+3}(x) - T'_{2k+3}(0) T_1(x)$. 

Furthermore, since we are solving a nonlinear eigenvalue problem, we considered the eigenvalue $\omega$ as an additional unknown in our nonlinear system of equations for the spectral coefficients, making it necessary to include an extra equation into the system. For the $\ell=0$ case we included the extra equation $F_{\ell=0}(r=0)=F_{0,0}$ with $F_{0,0}$ an arbitrary parameter (following~\cite{Grandclement2014:rotatingBSs}), while for $\ell>0$ we included the extra equation $\alpha(r=0)=\alpha_0$ with $\alpha_0$ arbitrary (as was done in~\cite{Alcubierre:2022}). These extra equations allow us to parameterize our numerical solutions with the values of $F_{0,0}$ or $\alpha_0$, respectively. Finally, we assessed the accuracy of our spectral code for the four cases of $\ell=0,1,2,3$ following the procedure presented in~\cite{Grandclement2014:rotatingBSs}. We confirmed the spectral convergence of our numerical solutions by verifying the exponential decay of the relative error in the frequency $\omega$, as well as the the difference between the ADM and Komar masses, as the order of the spectral decomposition was increased.  Since the compactification technique allows us to obtain values at infinity, the ADM mass and the Komar mass were calculated in our code using the expressions presented in~\cite{Grandclement2014:rotatingBSs,Gourgoulhon2012,gourgoulhon2011:introduction}: 
\begin{equation}\label{expr:ADM_mass-Komar_mass}
M_{\rm ADM} = - 2 \lim_{r\rightarrow\infty} \left( r^2 \: \frac{d\psi}{dr} \right) \; , \qquad
M_{\rm Komar} = \lim_{r\rightarrow\infty} \left( r^2 \: \frac{d\alpha}{dr} \right) \; .
\end{equation}


\subsection{Ground state configurations}

For simplicity, in this subsection and the next, all our solutions are obtained by setting the mass parameter of the Proca fields to $\mu=1$. This is possible because the solutions can be rescaled to arbitrary values of $\mu$ due to the invariance of the system of equations~\eqref{eq:initial_data:second_order:F}-\eqref{eq:initial_data:second_order:lapse} under the scaling transformation:
\begin{equation}
\mu \mapsto \lambda \mu \; , \quad
\omega \mapsto \lambda \omega \: , \quad
r \mapsto \lambda^{-1}r \; , \quad
b_\ell \mapsto \lambda^{-1} b_\ell \; ,
\end{equation}
with $(F_\ell,a_\ell,\alpha,\psi)$ unchanged.

\begin{figure}
\centering
\includegraphics[width=1.0\textwidth]{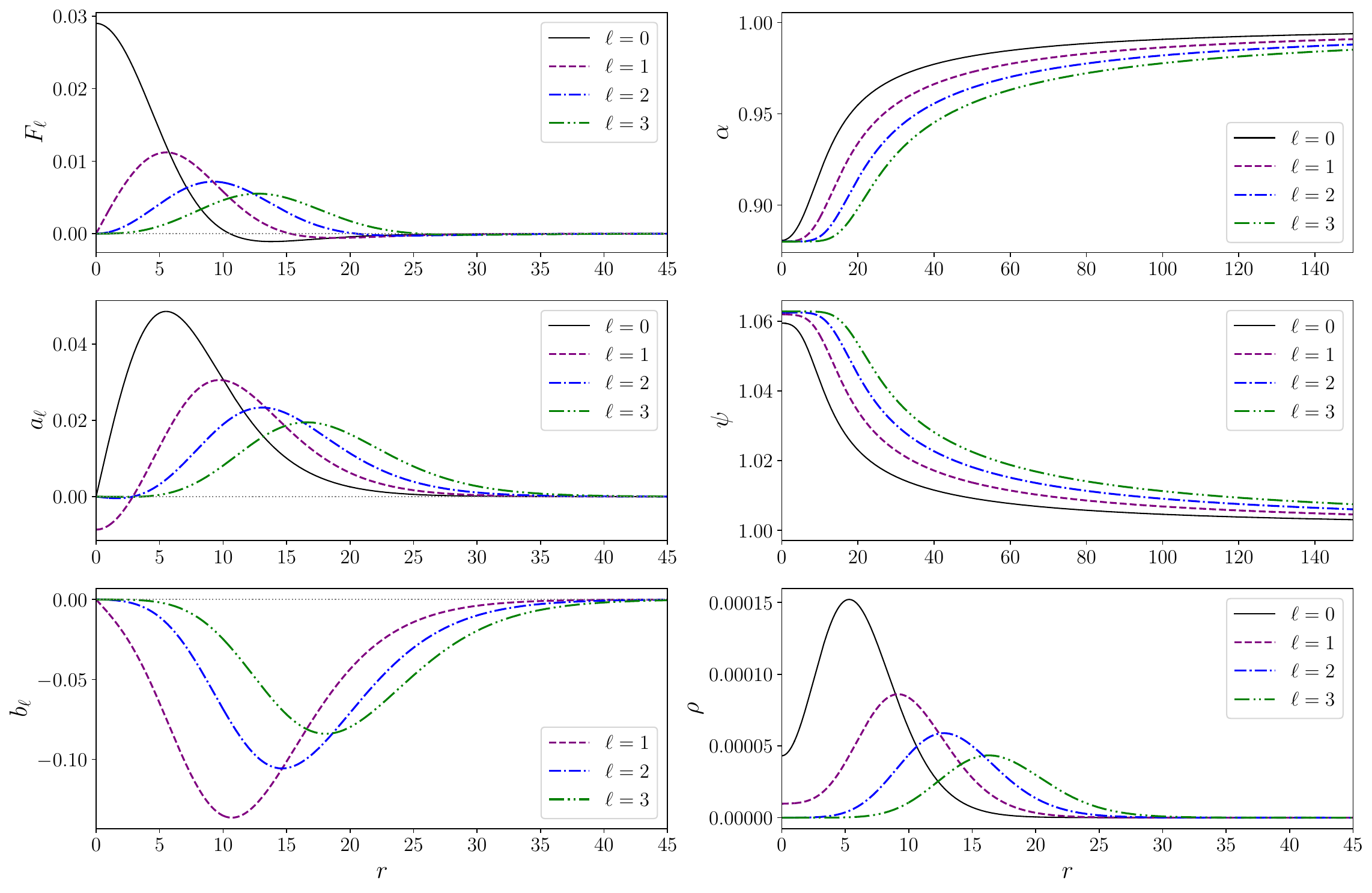}
\caption{Radial profiles of $\ell$-Proca stars ground state configurations for $\ell=0,1,2,3$ and $\alpha_0=0.88$. Left panel: matter functions $F_\ell(r), a_\ell(r)$ and $b_\ell(r)$. Right panel: metric functions $\alpha(r)$ and $\psi(r)$, and energy density $\rho(r)$. }
\label{fig:fundamental_solutions}
\end{figure}

In Figure~\ref{fig:fundamental_solutions} we show the radial profiles of the matter functions $(F_\ell,a_\ell,b_\ell)$ (left panel), the metric functions $\alpha$ and $\psi$ (top and middle plots of right panel), and the energy density $\rho$ (bottom plot of right panel) for a representative $\ell$-Proca star configuration for the first four values of $\ell$, and with the same value of the lapse at the origin $\alpha_0=0.88$. As we can observe, the radial profiles of $(F_0,a_0)$ and $\rho$ computed using our spectral code with $\ell=0$ (solid black line) reproduce the same type of profiles that were first obtained using a 1-dimensional shooting method in the original paper on Proca stars~\cite{BRITO2016291}.  Notice how in all cases the functions $(F_\ell,a_\ell,b_\ell,\rho)$ decay rapidly for large $r$ (see below), while the lapse $\alpha$ and conformal factor $\psi$ approach 1 (their asymptotic value for Minkowski spacetime) more slowly.

\begin{figure}
\centering
\includegraphics[width=1.0
\textwidth]{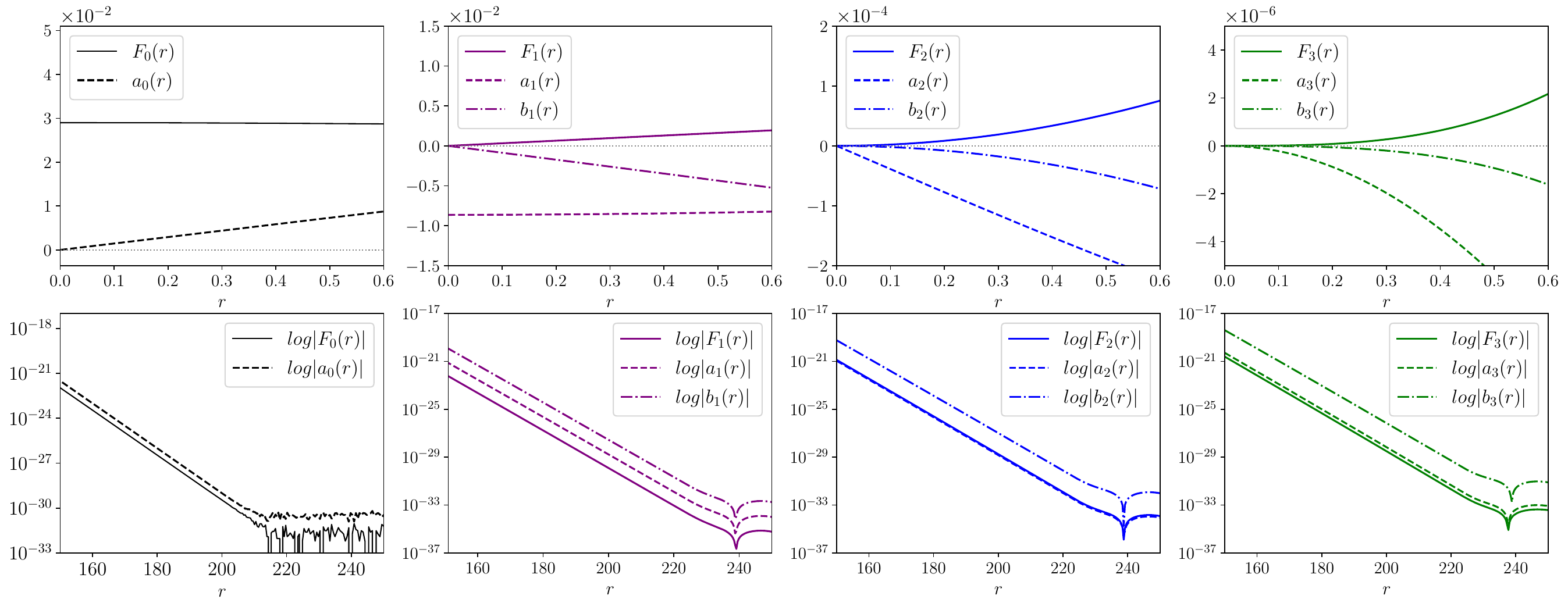}
\caption{Asymptotic behavior of the matter functions appearing in Fig.~\ref{fig:fundamental_solutions} in the neighborhood of the origin (top row), and for large $r$ (bottom row). The latter plots are presented in a logarithmic scale.}
\label{fig:boundary_behavior}
\end{figure}

In order to perform a detailed analysis of the radial profiles of $(F_\ell,a_\ell,b_\ell)$, especially for non-trivial cases of $\ell$, we make a closeup of them in Figure~\ref{fig:boundary_behavior} showing their asymptotic behavior for both small and large values of $r$. The plots in the top row show their behavior close the origin, while the bottom row plots show their behavior for large values of $r$ in a logarithmic scale. The numerically obtained radial profiles in both cases are consistent with the analytically predicted behaviors discussed in section~\ref{sec:localsolutions}, {\em i.e.}\/ with the two-free-parameter dependent local solutions near the origin~\eqref{F:local_solution}-\eqref{b:local_solution}, and the decaying exponential solutions far away as evidenced by the linear behavior in the logarithmic scale. Notice also that for very large values of $r$ the exponential decay stops and the functions become essentially constant (with some noise), this is caused by the unavoidable round off error of the numerical calculations.

About the two-free-parameter dependency near the origin, we deduce its presence in our numerical solutions by noticing the following: (i) In the top row of Fig.~\ref{fig:boundary_behavior} the radial dependencies follow the first dominant power of the power series solutions of~\eqref{F:local_solution}-\eqref{b:local_solution}, and since we observe different signs in the growth of $F_\ell \sim c_1 r^\ell$ and $ b_\ell \sim c_2 r^\ell$ with $\ell=1,2,3$, we can infer that their coefficients  have non-zero values with different signs, more explicitly $c_1>0$ and $c_2<0$. (ii) This inference is even more strongly supported by looking at the behavior of the function $a_\ell$ for $\ell \neq$ 0 (middle left panel plot in Fig.~\ref{fig:fundamental_solutions}), using its analytical local behavior close the origin given by~\eqref{a:local_solution}. For $\ell>0$ the function $a_\ell$ first becomes slightly negative as we move away from the origin and rapidly changes sign (it has a node, see below), which is explained by the fact that the coefficient of $r^{\ell-1}$ is negative, while the coefficient of $r^{l+1}$ is positive and rapidly dominates as we move away from the origin.  The results presented in Figures~\ref{fig:fundamental_solutions} and~\ref{fig:boundary_behavior} prove the effectiveness and sufficiency of our implemented regularity and boundary conditions. They also confirm the lack of regularity at the origin of the 1-Proca star configuration due to the non-zero value $a_{\ell=1}(r=0)=c_2$, as we analytically predicted in Section~\ref{sec:localsolutions} above.

\begin{figure}
\centering
\includegraphics[width=1.0
\textwidth]{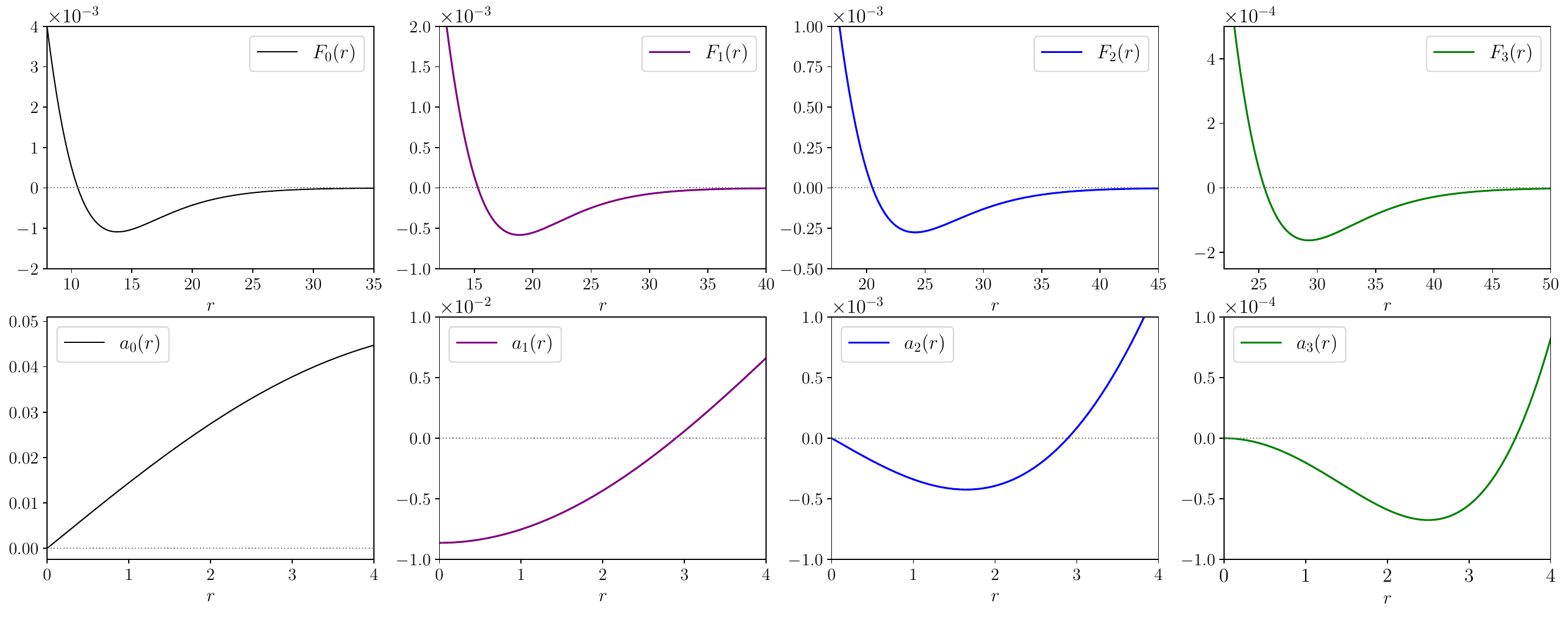}
\caption{Zoom of the node locations for the matter functions $F_\ell$ and $a_\ell$ appearing in Fig.~\ref{fig:fundamental_solutions}. Notice the different radial scales: the nodes for the functions $a_\ell$ are located very close to the origin.}
\label{fig:nodes_in_solutions}
\end{figure}

Observing Figure~\ref{fig:fundamental_solutions} more carefully reveals that the representative solutions shown here for both $F_\ell$ and $a_\ell$ present one node for non-trivial values of $\ell$. In order to show this more clearly we plot in Figure~\ref{fig:nodes_in_solutions} the region where these nodes are located. The top row of the Figure shows the nodes of $F_\ell(r)$ for the different values of $\ell$. The node in $F_0(r)$ was already reported in reference~\cite{BRITO2016291}, and our results show that this node is also present for the $\ell>0$ cases.  The bottom row shows the nodes of $a_\ell(r)$.  Notice in particular that the $\ell=0$ case is nodeless, which is consistent with the results of reference~\cite{BRITO2016291}, while the nodes that appear for $\ell>0$ constitute a new feature that $\ell$-Proca stars possess in contrast with the standard $\ell=0$ Proca star. On the other hand, for all our solutions the function $b_\ell(r)$ presents no nodes (see bottom left panel plot of Fig.~\ref{fig:fundamental_solutions}).  In order to ensure that the former solutions with nodes in $F_\ell$ and $a_\ell$ are the $\ell$-Proca stars ground state configurations, we used our spectral code to search for configurations with less energy for a given frequency $\omega$, without success.
We must therefore conclude that the solutions presented in Figure~\ref{fig:fundamental_solutions} do correspond to the ground state (minimum energy) configurations of the $\ell$-Proca stars.

Finally, the behavior of the matter functions close to the origin, as given by Eqs.~\eqref{matter:local:ell=0}-\eqref{matter:local:ell>0} and shown in Fig.~\ref{fig:fundamental_solutions}, implies that for the $\ell=2,3$ cases the energy density vanishes at the origin, which means that these configurations have a spherical shell morphology, similar to that of the $\ell$-boson stars presented in~\cite{Alcubierre:2018,Alcubierre:2019}. On the other hand, the energy density for the cases $\ell=0,1$ presents a non-zero local minimum at the center.


\subsection{Families of solutions}

\begin{figure}
\centering
\includegraphics[width=1.0
\textwidth]{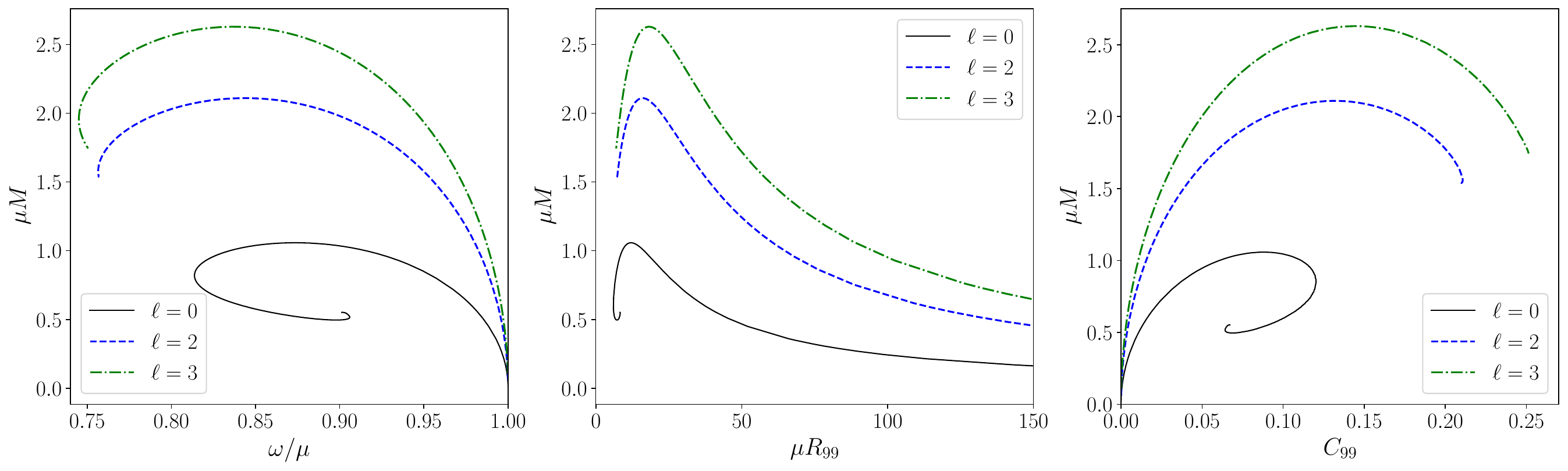}
\caption{Families of solution curves for ground state modes of $\ell$-Proca stars with $\ell=0,2,3$. Each point along these curves represents a specific solution with a given mass $M_{\rm ADM}$, frequency $\omega$, effective radius $R_{99}$, and effective compactness $C_{99}$.}
\label{fig:families_of_solutions}
\end{figure}

In this Section we characterize the families of $\ell$-Proca star ground state solutions with different angular momentum numbers $\ell=0,2,3$. Notice that the $\ell=1$ case has not been considered here since it is not regular at the center, and therefore does not represent an everywhere regular bosonic star solution. The characterization of the different families is based on the relation of the total ADM mass $M_{\rm ADM}$ of our solutions with: (i) the time frequency $\omega$, (ii) the effective radius $R_{99}$, and (iii) the effective compactness $C_{99}$. These relations are showed in the left, middle and right panels of Figure~\ref{fig:families_of_solutions}, respectively. The total ADM mass $M_{\rm ADM}$ is computed from the expression~\eqref{expr:ADM_mass-Komar_mass}. Notice that bosonic stars in general do not have a well defined boundary as the matter content decays exponentially for large $r$.  It is therefore standard practice to define an effective radius $R_{99}$ by taking it as the areal radius of the sphere which contains $99\%$ of the total mass~\cite{Alcubierre:2018, Alcubierre:2022}. For this calculation it was necessary to use the Misner-Sharp mass function, which in isotropic coordinates takes the form $m(r) = - 2 r^2 \psi' ( \psi + r \psi' )$, and  the areal radial coordinate  $r_a$ which in terms of the isotropic radius $r$ is given as $r_a = r \psi^2(r)$. Using these expressions we find the value of $r_a$ such that $m(r)=0.99 M_{\rm ADM}$.  To define the compactness of the $\ell$-Proca stars we use the definition $C_{99}:= M_{ADM}/R_{99}$.

The left panel of Fig.~\ref{fig:families_of_solutions} shows how the total mass $M_{\rm ADM}$ as a function of the frequency $\omega$ follows a spiral-like curve, common in standard boson and neutron star configurations.  As was observed for the case of $\ell$-boson stars in~\cite{Alcubierre:2018}, we also find more massive configurations as the value of $\ell$ grows, which is consistent with the fact that the rotational energy of the constituent spinning fields now forms part of the total energy. The middle panel of Fig.~\ref{fig:families_of_solutions} shows the dependence of the total mass with the effective radius $R_{99}$.  We can see that as the radius decreases from high values, the mass first increases monotonically up to a maximum value and then decreases again. Similarly, in the right panel of Fig.~\ref{fig:families_of_solutions} we see that the
effective compactness $C_{99}$ increases from zero up to a maximum value and then decreases.  Notice here that the solutions with maximum compactness do not correspond with the solutions with maximum mass. It is also clear that by increasing the angular momentum number $\ell$ one finds more compact objects. For the case with $\ell=3$ the maximum effective compactness reaches a value of $C_{99} \sim 0.25$, which should be compared with a value of $M/R=0.5$ for a Schwarzschild black hole. 

As is usual with other bosonic stars, the solution with maximum total mass divides the $\ell$-Proca star family of solutions into two branches: a branch with low values of compactness (right branch in the $M(\omega)$ curve and left branch in the $M(C_{99})$ curve), and a branch with highly compact configurations (left branch in the $M(\omega)$ curve and right branch in the $M(C_{99})$ curve). In order to see if these two branches correspond to stable and unstable configurations as in the case of standard boson stars one would need to do either dynamical simulations of our configurations or a linear stability analysis, but we will leave such considerations for a future work.

All the configurations presented here fulfill two conditions derived from the analytical solutions in the asymptotic regimes discussed in section~\ref{sec:localsolutions}. The first condition, $\omega > \alpha_0 \mu$, ensures real values for the coefficients $\kappa_1$ and $\kappa_2$~\eqref{expr:kappas}, that in turn guarantee a real domain for the Bessel functions needed to obtain the local solutions close the origin~\eqref{F:local_solution}-\eqref{b:local_solution}. The second condition, $\omega<\mu$, ensures that far away we have exponential solutions, as opposed to oscillating sinusoidal solutions.  Both these conditions can be summarized as $\alpha_0 < \omega / \mu < 1$. In table~\ref{table:max_mass configurations} we present the characteristic parameters for the configurations with maximum mass for the different values of $\ell$. Notice that, since we have set the mass parameter for the Proca fieldS to $\mu=1$, these configurations satisfy the two conditions just mentioned. 

\setlength{\tabcolsep}{8pt}
\begin{table}[h]
\centering
\begin{tabular}{c c c c c c}
\\ \hline
Angular momentum  & $\alpha_0$ & $\omega$  & $M_{max}$  & $R_{99}$ & $C_{99}$  \\ \hline\hline
$\ell=0$  &  0.733  & 0.874    & 1.058 & 12.029 & 0.088 \\
$\ell=2$  &  0.730  & 0.844    & 2.110 & 15.992 &  0.132 \\
$\ell=3$  &  0.728  & 0.836    & 2.629 & 18.184 &  0.146 \\ \hline
\end{tabular}
\caption{Values of the central lapse function $\alpha_0$, frequency $\omega$, ADM mass $M_{\rm ADM}$, effective radius $R_{99}$, and effective compactness $C_{99}$ (in Planck units), for the configurations with maximum total mass in the curves shown in Figure~\ref{fig:families_of_solutions}. Notice that all these configurations have been obtained by taking $\mu=1$.}
\label{table:max_mass configurations}
\end{table}


\section{Discussion and Conclusions}
\label{sec:CONCLUSIONS}

We have found a new bosonic star configuration that solves the spherically symmetric Einstein-(multi)Proca system. We call these new configurations $\ell$-Proca stars, where the parameter $\ell$ is a fixed integer value that corresponds to the azimuthal angular momentum number of all the available spherical harmonics that prescribe the angular dependency of the $2\ell+1$ constituent complex Proca fields. These solutions generalize the usual one-field spherical Proca stars ($\ell=0$), sharing features such as the staticity and spherical symmetry of the spacetime, the presence of one radial node in the scalar potential of its ground state configuration, and the fact that the family of solutions forms a spiral-like curve in the mass vs. frequency space.  Our solutions also include some new features when we consider non-trivial values of $\ell$: (i) the existence of non-zero angular components for the 3-vector potentials and ``electric'' fields, (ii) the existence of non-zero ``magnetic'' fields, and (iii) the appearance of one node in the radial component of the 3-vector potentials of the ground state. In this way we have enlarged the solution space for Proca stars introducing more complex configurations with a richer structure, that in a similar way to the $\ell$-boson stars of reference~\cite{Alcubierre:2018}, describe more massive and compact objects as we increase the value of $\ell$.

The radial profiles we obtained numerically (using spectral methods), and the imposed  harmonic time-dependency, express together the complete evolution of the ground state equilibrium configurations of our $\ell$-Proca stars. As a next step one should study the time evolution of these configurations under perturbations that break its dynamical equilibrium. Our numerical solutions give us a way to begin such studies since the computed radial profiles can be taken as initial data for such simulations. Indeed, some preliminary 3D numerical evolutions without imposing any spatial symmetries have already been performed in order to asses the stability of our solutions, and will be reported in a future paper.

It is important to stress that our results rule out the $\ell=1$ Proca stars as physical solutions since they present a loss of regularity due to the fact that the radial component of their constituent 3-vector potential does not vanish at the origin. This lack of regularity is similar to the one located at the origin in the case of the electrostatic potential of a point charge, except in the fact that it does not produce infinite values for the electric field or the energy density. In this sense it might correspond to a solution that has some kind of point-like source for the Proca fields.  Further studies should clarify this issue.

As a final comment, recent studies presented while we were writing this manuscript~\cite{herdeiro2023} seem to indicate that the standard spherical Proca stars $(\ell=0)$ are in fact excited state configurations that are unstable under non-spherically symmetric perturbations, losing their spherical symmetry and decaying to a prolate static ground state configuration without any nodes (despite being stable under perturbations in spherically symmetric numerical evolutions~\cite{BRITO2016291,Sanchis_Gual_2017}). In the light of these results it will be important to explore whether the $\ell$-Proca stars are truly ground state solutions or are also excited modes of non-spherically symmetric configurations. Once this is clarified, we could find out if the conjecture of the existence of a family of multi-field and multi-frequency Proca stars similar to that of the standard (scalar) boson stars~\cite{Sanchis-Gual:2021} still holds for non-trivial values of $\ell$.


\acknowledgments

We would like to thank Víctor Jaramillo for helping us clarify our doubts about spectral methods algorithms, and Olivier Sarbach and Nicolas Sanchis-Gual for interesting discussions and helpful comments. This work was partially supported by CONAHCyT Network Projects No. 376127 and No. 304001, and DGAPA-UNAM project IN100523. Claudio Lazarte also acknowledges support from a CONAHCyT National Graduate Grant and from the Generalitat Valenciana through a Santiago Grisolía
Grant (CIGRIS/2022/164).


\bibliographystyle{apsrev4-2}
\bibliography{referencias}


\end{document}